\documentclass[aps,prl,floatfix,twocolumn,tightenlines,amsmath,amssymb]{revtex4}

\usepackage{graphicx}
\usepackage{amssymb}
\usepackage{color}
\usepackage{psfrag}
\usepackage{ifsym}
\usepackage{epstopdf}
\usepackage{cancel}
\usepackage{dsfont}
\usepackage{hyperref}
\usepackage{bold-extra}

\newcommand{\ket}[1] {| #1 \rangle}
\newcommand{\braket}[2] {\langle #1 | #2 \rangle}
\newcommand{\ketbra}[1] {| #1 \rangle\!\langle #1 |}

\newcommand{\Tr} {\operatorname{Tr}}

\newcommand{\ve}{\varepsilon}
\newcommand{\sandwich}[3]{\mbox{$ \langle #1 | #2 | #3 \rangle $}}

\newcommand {\be} {\begin{eqnarray*}}
\newcommand {\ee} {\end{eqnarray*}}
\newcommand {\bea} {\begin{eqnarray}}
\newcommand {\eea} {\end{eqnarray}}

\newcommand{\bl}{{\bigl(}}
\newcommand{\br}{{\bigr)}}

\makeatletter
\renewcommand*\env@matrix[1][c]{\hskip -\arraycolsep
  \let\@ifnextchar\new@ifnextchar
  \array{*\c@MaxMatrixCols #1}}
\makeatother

 {\begin{list}{}%
         {\setlength{\leftmargin}{#1} \setlength{\rightmargin}{#2}}%
         \item[]%
 }
 {\end{list}}

\begin{document}
\title{Experimental Simulation of Closed Timelike Curves}

\author{Martin Ringbauer$^{1,2}$\footnote{Electronic address: {m.ringbauer@uq.edu.au}}, Matthew A. Broome$^{1,2}$, Casey R.~Myers$^{1}$, Andrew G. White$^{1,2}$ and Timothy C.~Ralph$^{2}$}
\affiliation{$^1$Centre for Engineered Quantum Systems, $^{2}$Centre for Quantum Computer and Communication Technology, School of Mathematics and Physics, University of Queensland, Brisbane,   QLD 4072, Australia}

\begin{abstract}
Closed timelike curves are among the most controversial features of modern physics. As legitimate solutions to Einstein's field equations, they allow for time travel, which instinctively seems paradoxical. However, in the quantum regime these paradoxes can be resolved leaving closed timelike curves consistent with relativity. The study of these systems therefore provides valuable insight into non-linearities and the emergence of causal structures in quantum mechanics---essential for any formulation of a quantum theory of gravity. Here we experimentally simulate the non-linear behaviour of a qubit interacting unitarily with an older version of itself, addressing some of the fascinating effects that arise in systems traversing a closed timelike curve. These include perfect discrimination of non-orthogonal states and, most intriguingly, the ability to distinguish nominally equivalent ways of preparing pure quantum states. Finally, we examine the dependence of these effects on the initial qubit state, the form of the unitary interaction, and the influence of decoherence.  
\end{abstract}

\maketitle

\section{Introduction}
\noindent One aspect of general relativity that has long intrigued physicists is the relative ease with which one can find solutions to Einstein's field equations that contain closed timelike curves (CTCs)---causal loops in space-time that return to the same point in space and time~\cite{PhysRevLett.61.1446, MorrisAmJPhys88, PhysRevLett.66.1126}. Driven by apparent inconsistencies---like the grandfather paradox---there have been numerous efforts, such as Novikov's self-consistency principle~\cite{NovikovBook83} to reconcile them or Hawking's chronology protection conjecture~\cite{hawking1992conjecture}, to disprove the existence of CTCs. While none of these classical hypotheses could be verified so far, the situation is particularly interesting in the quantum realm. In his seminal 1991 paper Deutsch showed for quantum systems traversing CTCs there always exist unique solutions, which do not allow superluminal signalling~\cite{deutsch1991quantum,ralph2010information}. Quantum mechanics therefore allows for causality violation without paradoxes whilst remaining consistent with relativity.

Advances in the field of Deutsch CTCs have shown some very surprising and counter-intuitive results, such as the solution of NP-complete problems in polynomial time~\cite{bacon2004quantum}, unambiguous discrimination of any set of non-orthogonal states~\cite{PhysRevLett.102.210402}, perfect universal quantum state cloning~\cite{PhysRevA.88.022332,Brun2013} and the violation of Heisenberg's uncertainty principle~\cite{pienaar2013open}. The extraordinary claims of what one could achieve given access to a quantum system traversing a CTC have been disputed in the literature, with critics pointing out apparent inconsistencies in the theory such as the information paradox or the linearity trap~\cite{Bennet1992linear, PhysRevA.84.056301}. However, it has been shown that the theory can be formulated in such a way that these inconsistencies are resolved ~\cite{ralph2010information, PhysRevA.84.056302}. 

Modern experimental quantum simulation allows one to ask meaningful questions that provide insights into the behaviour of complex quantum systems.
Initial results have been obtained in various areas of quantum mechanics~\cite{KitagawaNatComm3,MaNatPhys7,Simon2011} and in particular in the field of relativistic quantum information~\cite{Gerritsma2010,Casanova2011,Philbin2008,Menicucci2010,Lloyd2011a}. 
This recent experimental success, coupled with the growing interest for the study of non-linear extensions to quantum mechanics, motivates the question of whether the fundamentally non-linear dynamics and the unique behaviour arising from CTCs can be simulated experimentally.

In this article we use photonic systems to simulate the quantum evolution through a Deutsch CTC. We demonstrate how the CTC-traversing qubit adapts to changes in the input state $\ket{\psi}$, and unitary interaction $U$ to ensure physical consistency according to Deutsch's consistency relation~\cite{deutsch1991quantum}. We observe non-linear evolution in the circuit suggested by Bacon~\cite{bacon2004quantum} and enhanced distinguishability of two non-orthogonal states after the action of an optimised version of a circuit proposed by Brun et al.~\cite{PhysRevLett.102.210402}. Using the self-consistent formulation of Ref.~\cite{ralph2010information} we then move beyond the simplest implementations and find a striking difference in the behaviour of the system for direct as opposed to entanglement-assisted state preparation. Finally, we explore the system's sensitivity to decoherence. 

\begin{figure}[h!]
\begin{center}
\includegraphics[width=0.95\columnwidth]{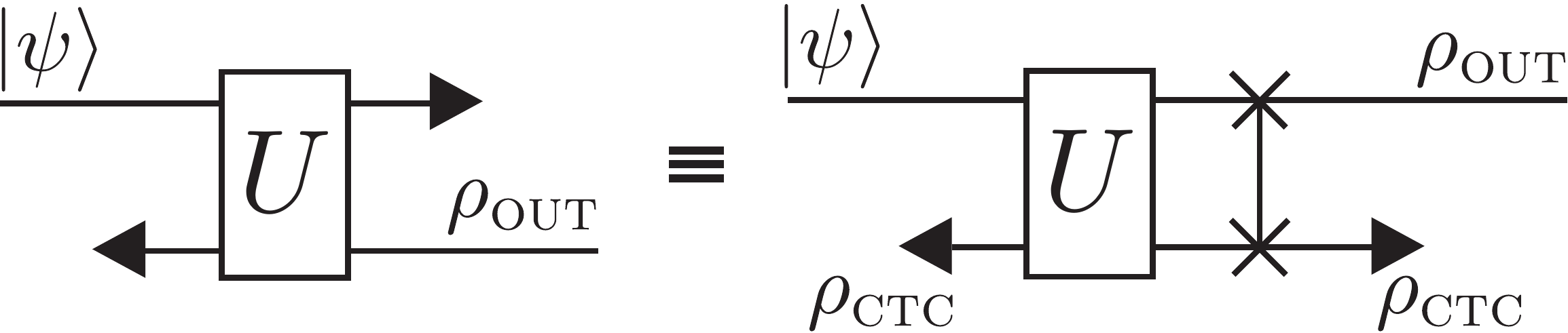}
\vspace{-5mm}
\end{center}
\caption{\textbf{Model of a quantum state $\ket\psi$ interacting with an older version of itself.} This situation can equivalently be interpreted as a chronology-respecting qubit interacting with a qubit trapped in a CTC. The CTC in general consists of a causal worldline with its past and future ends connected via a wormhole (indicated by black triangles).}
\label{fig:Deutsch}
\end{figure}
\vspace{-5mm}

\section{Results}
\noindent\textbf{The Deutsch model.}
While there has been some recent success on alternative models of CTCs based on postselection~\cite{Lloyd2011,Brun2011,Lloyd2011a}, we focus on the most prominent model for describing quantum mechanics in the presence of CTCs, introduced by Deutsch~\cite{deutsch1991quantum}. Here a quantum state $\ket\psi$ interacts unitarily with an older version of itself, Fig.~\ref{fig:Deutsch}. With the inclusion of an additional \textsc{swap} gate, this can equivalently be treated as a two-qubit system, where a chronology-respecting qubit interacts with a qubit $\rho_{\textsc{ctc}}$ trapped in a closed timelike curve. The quantum state of $\rho_{\textsc{ctc}}$ in this picture is determined by Deutsch's consistency relation: 
\begin{equation}
\rho_{\textsc{ctc}} = \Tr_{1}\left[ U'\left(\ketbra\psi\otimes \rho_{\textsc{ctc}}\right)U'^{\dagger}\right] ,
\label{eq:Deutsch}
\end{equation}
where $U'$ is the unitary $U$ followed by a \textsc{swap} gate, Fig.~\ref{fig:Deutsch}. This condition ensures physical consistency---in the sense that the quantum state may not change inside the wormhole---and gives rise to the non-linear evolution of the quantum state $\ket\psi$. The state after this evolution is consequently given by $\rho_{\textsc{out}} {=} \Tr_{2}\left[ U'\left(\ketbra\psi\otimes \rho_{\textsc{ctc}}\right)U'^{\dagger}\right]$. The illustration in Fig.~\ref{fig:Deutsch} further shows that the requirement of physical consistency forces $\rho_{\textsc{ctc}}$ to adapt instantly to any changes in the surroundings, such as a different interaction unitary $U$ or input state $\ket\psi$. While Eq.~\eqref{eq:Deutsch} is formulated in terms of a pure input state $\ket\psi$ it can be directly generalised to mixed inputs~\cite{ralph2010information}.

\begin{figure}[h!]
\begin{center}
\includegraphics[width=1\columnwidth]{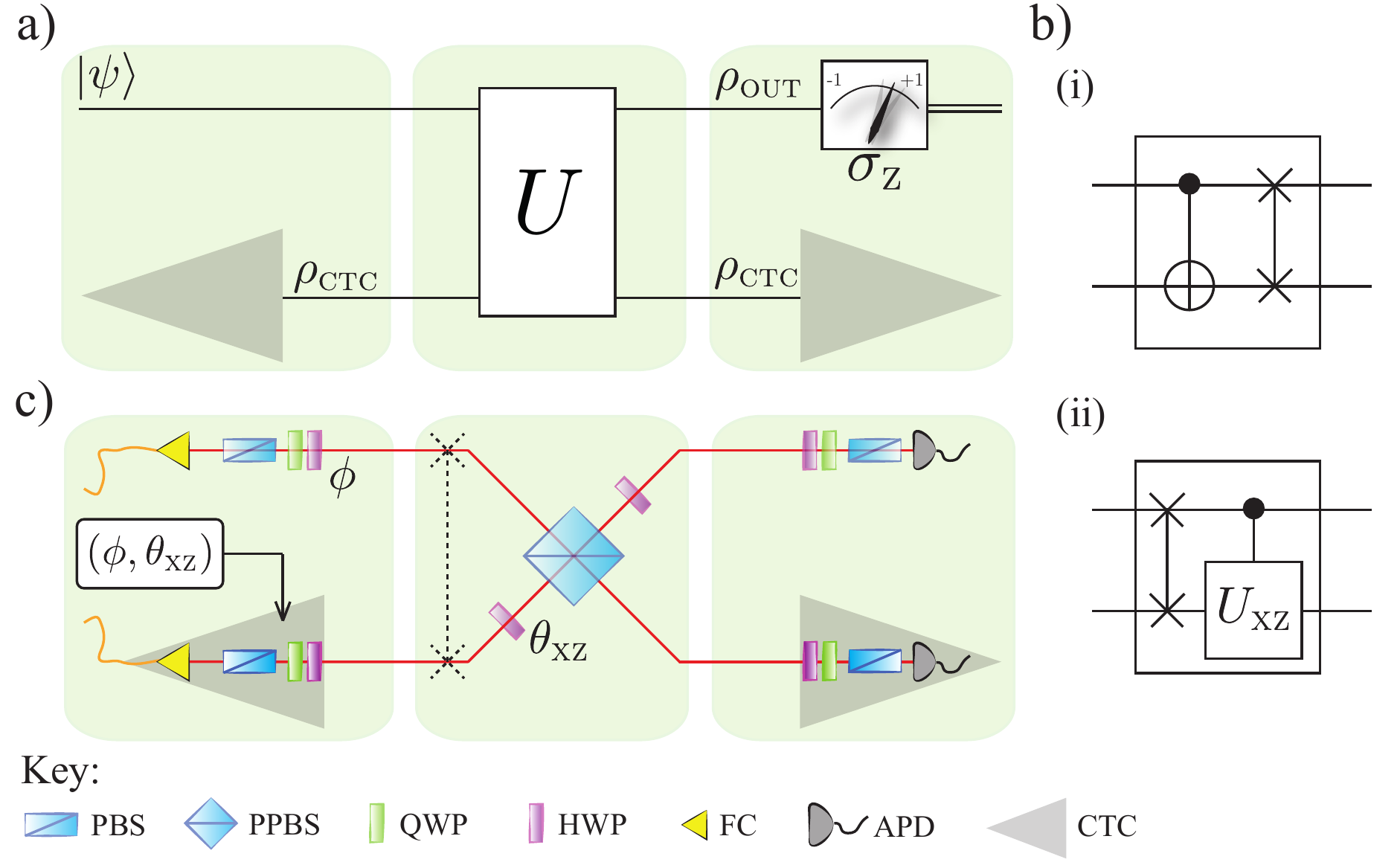}
\vspace{-7mm}
\end{center}
\caption{\textbf{Experimental details.} \textbf{a)} The circuit diagram for a general unitary interaction $U$ between the state $\ket{\psi}$ and the CTC system. \textbf{b)} The specific choice of unitary in the demonstration of the \textbf{(i)} non-linear evolution and \textbf{(ii)} perfect discrimination of non-orthogonal states. \textbf{c)} Experimental setup for case (ii). Two single photons, generated via spontaneous parametric down-conversion in a nonlinear $\beta$-barium-borate crystal, are coupled into two optical fibres (FC) and injected into the optical circuit. Arbitrary polarisation states are prepared using a Glan-Taylor polariser (POL), a quarter-wave (QWP) and a half wave-plate (HWP). Non-classical interference occurs at the central partially-polarising beam-splitter (PPBS) with reflectivities $\eta_{H}{=}0$ and $\eta_{V}{=}2/3$. Two avalanche photo-diodes (APD) detect the single photons at the outputs. The states $\ket{\psi}$ are chosen in the $\textsc{xz}$-plane of the Bloch sphere, parametrised by $\phi$ and $CU_\textsc{xz}$ is the corresponding controlled unitary, characterised by the angle $\theta_\textsc{xz}$. The $\textsc{swap}$ gate was realized via relabeling of the input modes.}
\label{fig:setup}
\end{figure}

$ $\\
\noindent\textbf{Simulating CTCs.}
Our experimental simulation of a qubit in the (pure) state $\ket{\psi}$ traversing a CTC relies on the circuit diagram shown in Fig.~\ref{fig:setup}a). A combination of single qubit unitary gates before and after a controlled-Z gate allows for the implementation of a large set of controlled-unitary gates $U$. Using polarisation-encoded single photons, arbitrary single qubit unitaries can be realised using a combination of quarter-wave (QWP) and half-wave plates (HWP); additional $\textsc{swap}$ gates before or after $U$ are implemented as a physical mode-swap.
The controlled-Z gate is based on non-classical (Hong-Ou-Mandel) interference of two single photons at a single partially polarising beam-splitter (PPBS) that has different transmittivities $\eta_{V}{=}1/3$ for vertical (V) and $\eta_{H}{=}1$ for horizontal (H) polarisation~\cite{Ralph2002}---a more detailed description of the implementation of the gate can be found in Ref.~\cite{Langford:CZ}. Conditioned on post-selection it induces a $\pi$ phase-shift when the two interfering single-photon modes are vertically polarised, such that $\ket{VV} \rightarrow -\ket{VV}$ with respect to all other input states.

One of the key features of a CTC is the inherently non-linear evolution that a quantum state $\ket\psi$ undergoes when traversing it. This is a result of Deutsch's consistency relation, which makes $\rho_\textsc{ctc}$ dependent on the input state $\ket\psi$. In order to simulate this non-linear behavior using linear quantum mechanics we make use of the effective non-linearity obtained from feeding extra information into the system. In our case we use the classical information about the preparation of the state $\ket\psi$ and the unitary $U$ to prepare the CTC qubit in the appropriate state $\rho_\textsc{CTC}$ as required by the consistency relation Eq.~\eqref{eq:Deutsch}. After the evolution we perform full quantum state tomography on the CTC qubit in order to verify that the consistency relation is satisfied.

\begin{figure}[h!]
\begin{center}
\includegraphics[width=1\columnwidth]{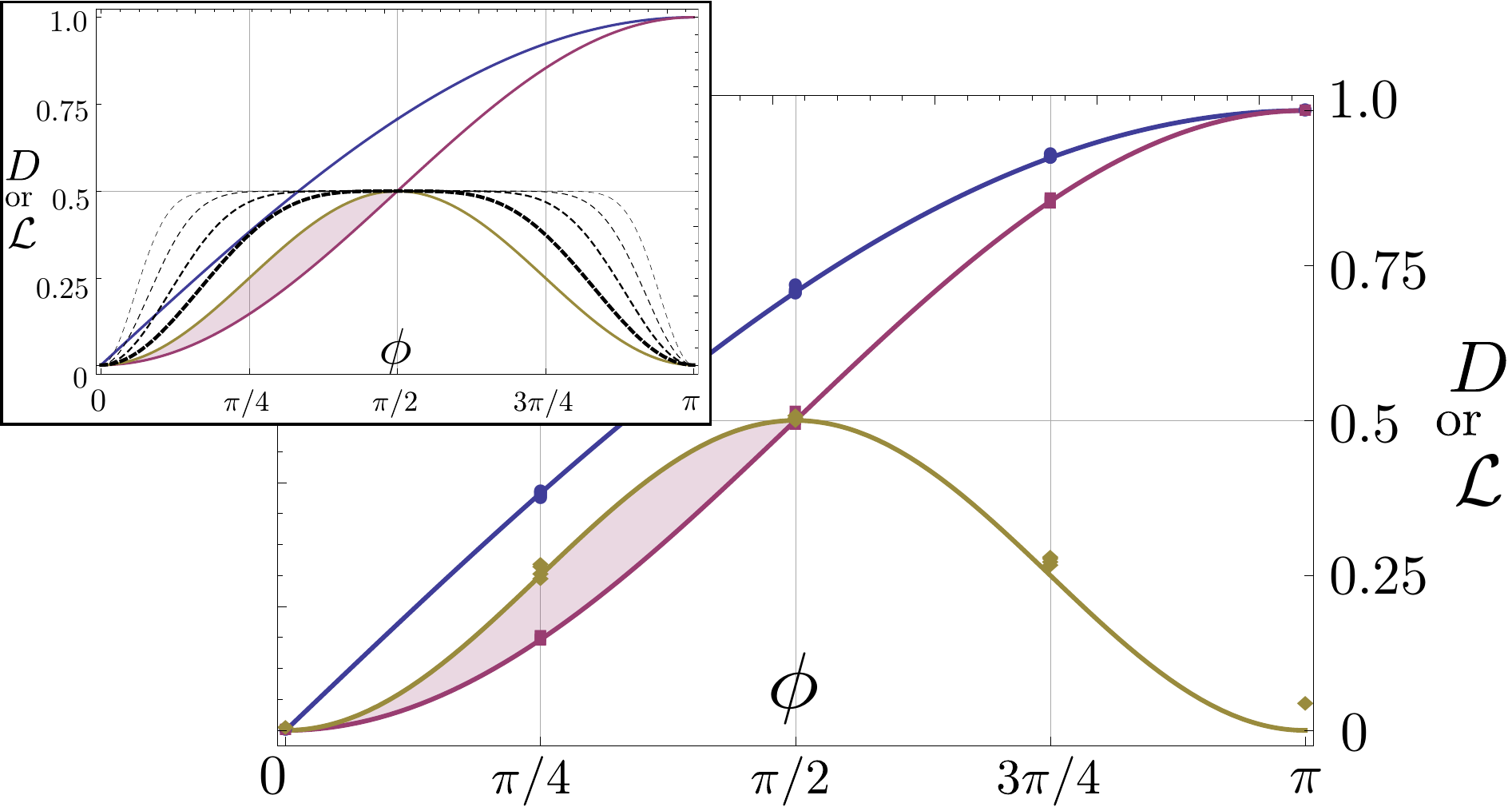}
\vspace{-7mm}
\end{center}
\caption{\textbf{Non-linear evolution in a Deutsch CTC with \textsc{swap}.\textsc{cnot} interaction.} Both the trace distance $\mathcal D$, and the $\sigma_z$-based distinguishability measure $\mathcal{L}$ (equal to within experimental error in this case) of the evolved states $\rho_\textsc{out}$ after the interaction with the CTC are shown as yellow diamonds. The blue circles (red squares) represent the measure $\mathcal D$ ($\mathcal{L}$) between the input states $\ket\psi$ and $\ket{H}$ in the case of standard quantum mechanics. Note that due to the phase-independence of the evolution in Eq.~\eqref{eq:nonlinear_Bacon} states that only differ by a phase collapse to a single data point. Crucially, the metric $\mathcal D$ does not capture the effect of the non-linearity, while $\mathcal{L}$ does, indicated by the red shaded region. Error bars obtained from a Monte Carlo routine simulating the Poissonian counting statistics are too small to be visible on the scale of this plot.
\textbf{Inset:} The dashed black lines with decreasing thickness represent theoretical expectations for $\mathcal D$ and $\mathcal L$ from $2,3,4$ and $5$ iterations of the circuit.}
\label{fig:results_nonlinear_bacon}
\end{figure}

$ $\\
\noindent\textbf{Non-linear evolution.}
As a first experiment we investigate the non-linearity by considering a Deutsch CTC with a \textsc{cnot} interaction followed by a $\textsc{swap}$ gate as illustrated in Fig.~\ref{fig:setup}b)(i). This circuit is well-known for the specific form of non-linear evolution:
\begin{equation}
\alpha\ket{H}+e^{i\varphi}\beta\ket{V}\to (\alpha^{4}+\beta^{4})\ketbra{H}+2\alpha^{2}\beta^{2}\ketbra{V} ,
\label{eq:nonlinear_Bacon}
\end{equation}
which has been shown to have important implications for complexity theory, allowing for the solution of NP-complete problems with polynomial resources~\cite{bacon2004quantum}. According to Deutsch's consistency relation, Eq.~\eqref{eq:Deutsch} the state of the CTC-qubit for this interaction is given by
\begin{equation}
\rho_{\textsc{ctc}}= \alpha^2 \ketbra{H} + \beta^2 \ketbra{V} .
\label{eq:nonlinear_Bacon_CTCstate}
\end{equation}
We investigate the non-linear behaviour  experimentally for 14 different quantum states $\ket{\psi}{=}\cos(\frac\phi 2)\ket{H}+e^{i\varphi}\sin(\frac\phi 2)\ket{V}$, with $\phi\in\{0,\frac{\pi}{4},\frac{\pi}{2},\frac{3\pi}{4},\pi\}$ and a variety of phases $\varphi\in\{0,2\pi\}$, where the locally available information $\phi$ and $\varphi$ is used to prepare $\rho_\textsc{ctc}$. In standard (linear) quantum mechanics no unitary evolution can introduce additional distinguishability between quantum states. To illustrate the non-linearity in the system we employ two different distinguishability measures: the trace-distance $\mathcal D(\rho_1,\rho_2){=}\frac 1 2 \Tr[ |\rho_1-\rho_2| ]$, where $|\rho |{=}\sqrt{\rho^\dagger \rho}$ and a single projective measurement with outcomes ``$+$'' and ``$-$'':
\begin{equation}
\mathcal{L}(\rho_1,\rho_2) = \sandwich{+}{\rho_1}{+}\sandwich{-}{\rho_2}{-}+\sandwich{-}{\rho_1}{-}\sandwich{+}{\rho_2}{+} .
\label{eq:sigmaZ}
\end{equation}
While the metric $\mathcal D$ is a commonly used distance measure it does not have an operational interpretation and requires full quantum state tomography in order to be calculated experimentally. The measure $\mathcal L$ in contrast is easily understood as the probability of obtaining different outcomes in minimum-error discrimination of the two states using a single projective measurement on each system. The operational interpretation and significance of $\mathcal L$ is discussed in more detail in the Supplemental Material. Both $\mathcal D$ and $\mathcal L$ are calculated between the state $\ket{\psi}$ and the fixed reference state $\ket H$ after being evolved through the circuit shown in Fig.~\ref{fig:setup}b)(i). The results are plotted and compared to standard quantum mechanics in Fig.~\ref{fig:results_nonlinear_bacon}. If the state $\ket{\psi}$ is not known then, based only on the knowledge of the reference state $\ket{H}$ and the evolution in Eq.~\eqref{eq:nonlinear_Bacon} it is natural and optimal to use the measure $\mathcal L$ with a $\sigma_{\textsc{z}}$-measurement.

We observe enhanced distinguishability for all states with an initial trace-distance to $\ket{H}$ smaller than $1/\sqrt{2}$ (i.e.\ $\phi{\leq}\frac\pi 2$), as clearly demonstrated by the $\sigma_z$-based measure, see Fig.~\ref{fig:results_nonlinear_bacon}. Note, however, that this advantage over standard quantum mechanics is not captured by the metric $\mathcal D(\rho_1,\rho_2)$ unless the non-linearity is amplified by iterating the circuit on the respective output at least 3 times, see inset of Fig.~\ref{fig:results_nonlinear_bacon}. This shows that the non-linearity is not directly related to the distance between two quantum states. By testing states with various polar angles for each azimuthal angle on the Bloch sphere, we confirm that any phase information is erased during the evolution and that the evolved state $\rho_\textsc{out}$ is indeed independent of $\varphi$, up to experimental error.
We further confirm, with an average quantum state fidelity of $\mathcal{F}=0.998(2)$ between the input and output state of $\rho_{\textsc{ctc}}$ in Eq.~\eqref{eq:nonlinear_Bacon_CTCstate}, that the consistency relation~\eqref{eq:Deutsch} is satisfied for all tested scenarios.

$ $\\
\noindent\textbf{Non-orthogonal state discrimination.}
While it is the crucial feature, non-linear state evolution is not unique to the \textsc{swap}.\textsc{cnot} interaction, but rather a central property of all non-trivial CTC interactions. Similar circuits have been found to allow for perfect distinguishability of non-orthogonal quantum states~\cite{PhysRevLett.102.210402}, leading to discomforting possibilities such as breaking of quantum cryptography~\cite{PhysRevLett.102.210402}, perfect cloning of quantum states~\cite{PhysRevA.88.022332,Brun2013}, and violation of Heisenberg's uncertainty principle~\cite{pienaar2013open}.
In particular it has been shown that a set $\{\ket{\psi_j}\}_{j=0}^{N-1}$ of $N$ distinct quantum states in a space of dimension $N$ can be perfectly distinguished using an $N$-dimensional CTC-system. The algorithm proposed by Brun et al.~\cite{PhysRevLett.102.210402} relies on an initial $\textsc{swap}$ operation between the input and the CTC-system, followed by a series of $N$ controlled unitary operations, transforming the input states to an orthogonal set, which can then be distinguished.

\begin{figure}[h!]
\begin{center}
\includegraphics[width=1\columnwidth]{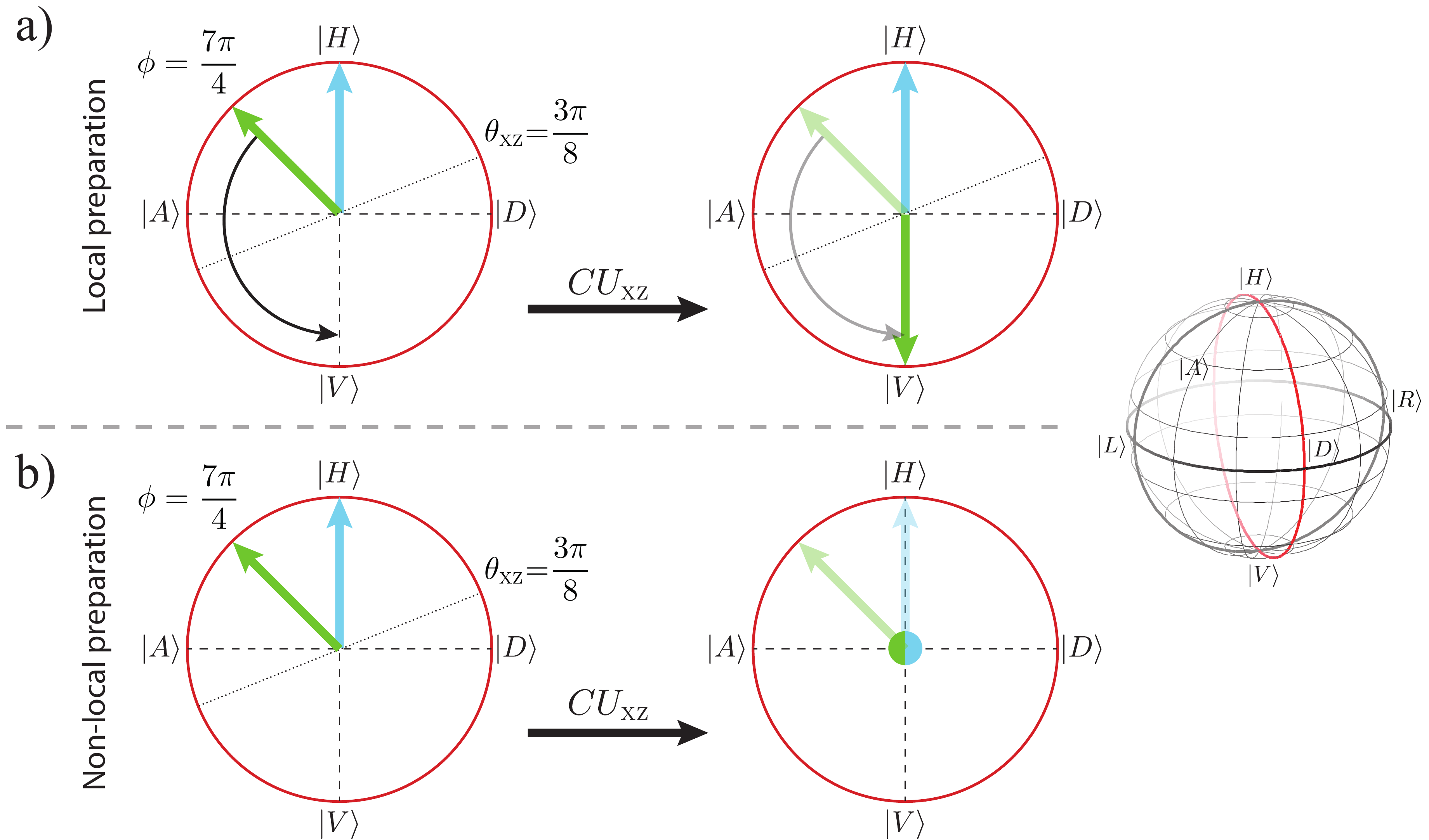}
\end{center}
\vspace{-5mm}
\caption{\textbf{Bloch-sphere evolution of states traversing a CTC.} In the case of \textbf{(a)} local state preparation, the state $\ket{\psi_0}{=}\ket{H}$ (blue) is unaffected by $CU_\textsc{xz}$, while $\ket{\psi_1}$ (green) undergoes a $\pi$ rotation about the axis defined by $\theta_\textsc{xz}$. The axis is chosen as $\theta_\textsc{xz}=\frac{\phi-\pi}{2}$ such that $\ket{\psi_1}\mapsto\ket{V}$ which can then be perfectly distinguished from $\ket{\psi_0}$. \textbf{(b)} For non-local preparation of the initial states and the same choice of $\theta_\textsc{xz}$ the controlled unitary maps both initial states to the maximally mixed state $\frac 1 2 \bl\ketbra{H}+\ketbra{V}\br$. The probability of distinguishing the two states is therefore $1/2$---as good as randomly guessing.}
\label{fig:state_evo}
\end{figure}

In our simulation of this effect we consider the qubit case $N{=}2$, which consequently would require two controlled unitary operations between the input state and the CTC system. We note, however, that without loss of generality the set of states to be discriminated can be rotated to the \textsc{xz}-plane of the Bloch sphere, such that $\ket{\psi_0}{=}\ket{H}$ and $\ket{\psi_1}{=}\cos(\frac{\phi}{2})\ket{H}{+}\sin(\frac{\phi}{2})\ket{V}$ for some angle $\phi$. In this case, the first controlled unitary is the identity operation $\mathcal{I}$, while the second performs a controlled rotation of $\ket{\psi_1}$ to $\ket{V}$ as illustrated in Fig.~\ref{fig:state_evo}a). In detail, the gate $CU_{\textsc{xz}}$ applies a $\pi$ rotation to the target qubit conditional on the state of the control qubit, about an axis in the $\textsc{xz}$-plane defined by the angle $\theta_{\textsc{xz}}$. For the optimal choice $\theta_{\textsc{xz}}=\frac{\phi-\pi}{2}$ the gate rotates the state $\ket{\psi_1}$ to $\ket{V}$, orthogonal to $\ket{\psi_0}$, enabling perfect distinguishability by means of a projective $\sigma_z$ measurement, see Fig.~\ref{fig:state_evo}a).

In practice the gate $CU_{\textsc{xz}}$ is decomposed into a controlled-Z gate between appropriate single qubit rotations, defining the axis $\theta_\textsc{xz}$. The latter are realised by half-wave plates before and after the PPBS, set to an angle of $\theta_{\textsc{xz}}/4$ with respect to their optic axis, see Fig.~\ref{fig:setup}c):
\begin{align}
CU_\textsc{xz} (\theta_{\textsc{xz}})&=(\mathcal{I} \otimes \operatorname{HWP}(\theta_{\textsc{xz}}/4))\cdot\textsc{cz}\cdot(\mathcal{I}\otimes \operatorname{HWP}(\theta_{\textsc{xz}}/4)) \nonumber \\[0.5em]
&=\begin{pmatrix}
1 & 0 & 0 & 0 \\
0 & 1 & 0 & 0 \\
0 & 0 & \cos(\theta_{\textsc{xz}}) & \sin(\theta_{\textsc{xz}}) \\
0 & 0 & \sin(\theta_{\textsc{xz}}) & -\cos(\theta_{\textsc{xz}})
\end{pmatrix} .
\label{eq:CUgate}
\end{align}

Note that relation $\eqref{eq:Deutsch}$ requires that $\rho_{\textsc{ctc}}{=}\ketbra{H}$, whenever the input state is $\ket{H}$, independent of the gate $CU_{\textsc{xz}}$.
Crucially, this consistency relation ensures that any physical CTC-system adapts instantly to changes in $\phi$ and $\theta_\textsc{xz}$, parametrising the input state and gate, respectively. In our simulation these two parameters are used to prepare the corresponding state $\rho_{\textsc{ctc}}$, as shown in Fig.~\ref{fig:setup}c). 

In a valid experimental simulation the input and output states $\rho_\textsc{ctc}$ have to match, i.e.\ $\rho_\textsc{ctc}$ has to satisfy relation~\eqref{eq:Deutsch}. This has been verified for all following experiments with an average quantum state fidelity of $\mathcal{F}{=}0.996(7)$.

In the experiment, we prepared near-pure quantum states directly on single photons using a Glan-Taylor polariser followed by a combination of a HWP and a QWP. We simulated CTC-aided perfect discrimination of non-orthogonal states for $32$ distinct quantum states $\ket{\psi_1}$ with $\phi\in [0,2\pi)$. For each state we implemented $CU_\textsc{xz}$ with the optimal choice of $\theta_{\textsc{xz}}{=}\frac{\phi-\pi}{2}$. Furthermore we tested the ability of this system to distinguish the set $\{\ket{\psi_0},\ket{\psi_1}\}$ given non-optimal combinations of $\phi$ and $\theta_\textsc{xz}$. For this we either chose $\phi{=}3\pi/2$ and varied the gate over the full range of $\theta_\textsc{xz}\in[-\frac\pi 2 , \frac\pi 2)$, or chose $CU_\textsc{xz}$ as a controlled Hadamard (optimal for $\phi{=}3\pi/2$) and varied the state $\ket{\psi_1}$ over the full range of $\phi\in[0,2\pi)$. The output state is characterized by quantum state tomography, which provides sufficient data to obtain $\mathcal L$ for arbitrary measurement directions as well as for the calculation of the trace-distance.

\begin{figure*}
\begin{center}
\includegraphics[width=1\textwidth]{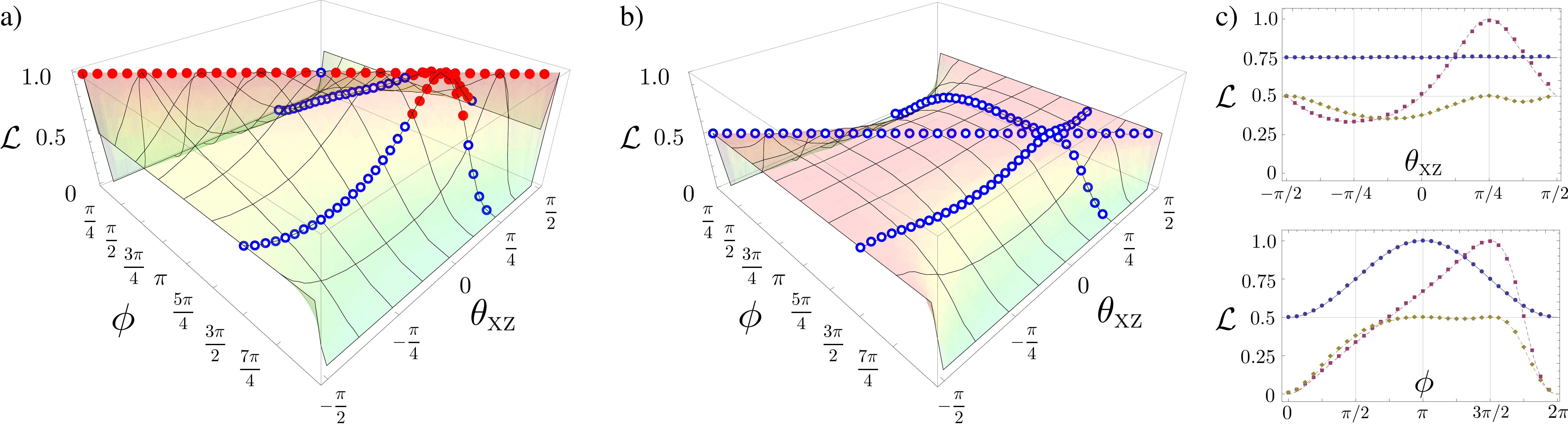}
\end{center}
\vspace{-5mm}
\caption{\textbf{Experimental results.} Probability of state discrimination for \textbf{a)} locally prepared and \textbf{b)} non-locally prepared states $\ket{\psi_0}{=}\ket{H}$ and $\ket{\psi_1}=\cos (\frac\phi 2)\ket{H}+\sin(\frac\phi 2)\ket{V}$ as measured by $\mathcal{L}$. The surface represents the theoretically predicted probability depending on the state and gate parameters $\phi$ and $\theta_\textsc{xz}$, respectively. Solid, red (open, blue) data-points indicate better (worse) performance than standard quantum mechanics. \textbf{c)} Cross-sectional views of the combined plots a) and b) reveal the rich structure in the dependencies on the initial parameters for \textbf{(top)} a fixed state ($\phi{=}3\pi/2$) and \textbf{(bottom)} a fixed gate ($\theta_\textsc{xz}{=}\pi/4$). Here red squares (yellow diamonds) correspond to the CTC case with local (non-local) preparation and blue circles represents standard quantum mechanics. Error bars obtained from a Monte Carlo routine simulating the Poissonian counting statistics are too small to be visible on the scale of this plot.}
\label{fig:results_dist}
\end{figure*}

Figure~\ref{fig:results_dist}a) illustrates the observed distinguishability $\mathcal{L}$ for the above experiments and compares it to the expectation from standard quantum mechanics. In the latter case the measure $\mathcal L$ is maximized by choosing the optimal projective measurement, based on the available information about the states $\ket{\psi_0}$ and $\ket{\psi_1}$. Crucially, the optimized $\mathcal L$ is directly related to the trace-distance $\mathcal D$ as $\mathcal L = \frac 1 2 (1+\mathcal{D}^2)$ and therefore captures the same qualitative picture, without the requirement for full quantum state tomography.
In the CTC case a $\sigma_{\textsc{z}}$-measurement is chosen, which is optimal when $\theta_{\textsc{xz}}=\frac{\phi-\pi}{2}$. Otherwise further optimization is possible based on the knowledge of $\theta_\textsc{xz}$ (see Supplemental Material and Fig.~\ref{fig:SuppResultsFullOpt} for more details). Furthermore, we note that the above scenario can also be interpreted in a state-identification rather than state-discrimination picture, which is discussed in more detail in the Supplemental Material and illustrated in Fig.~\ref{fig:SuppResultsIdentification}.

$ $\\
\noindent\textbf{Local vs.\ non-local state preparation.}
Due to the inherent non-linearity in our simulated system, care must be taken when describing mixed input states $\rho_{in}$. In particular a distinction between proper and improper mixtures can arise which is unobservable in standard (linear) quantum mechanics~\cite{Espagnat1976}. This ambiguity is resolved in Ref.~\cite{ralph2010information} by requiring the consistency condition to act shot-by-shot---i.e.\ independently in every run of the experiment---on the reduced density operator of the input state. For proper mixtures this means that $\rho_{in}$ is always taken as a pure state, albeit a different one shot-by-shot. For improper mixtures in contrast, $\rho_{in}$ will always be mixed.
A similar, but much more subtle and fascinating feature, which has received less attention in the literature so far occurs with respect to preparation of pure states~\cite{Cavalcanti2012}. While in standard quantum mechanics a pure state prepared directly (locally) on a single qubit is equivalent to one that has been prepared non-locally through space-like separated post-selection of an entangled resource state, this is not true under the influence of a CTC. The origin of this effect is not the non-linear evolution, but rather the local absence of classical information about the post-selection outcome. The role of locally available classical information in entanglement-based preparation schemes is a matter of current debate and still to be clarified.

A possible resource state for alternatively preparing $\ket{\psi_0}$ and $\ket{\psi_1}$ could be of the form $\ket{\Psi}{=}\frac 1 {\sqrt{2}} \bl {\ket{0}\otimes\ket{\psi_0}} + {\ket{1}\otimes\ket{\psi_1}} \br$, where projection of the first qubit onto the state $\ket{0}$ and $\ket{1}$ leaves the second qubit in the state $\ket{\psi_0}$ and $\ket{\psi_1}$, respectively. From the point of view of $\rho_{\textsc{ctc}}$, however, there exists no information about the outcome of this projective measurement. Hence it ``sees'' and adapts to the mixed state $\rho_{in}{=}\Tr_1 [ \ketbra{\Psi} ]{=}\frac 1 2 (\ketbra{\psi_0}{+}\ketbra{\psi_1})$. The state of the CTC qubit is therefore different for local and non-local preparation. If this was not the case, it would enable superluminal signalling, which is in conflict with relativity~\cite{Cavalcanti2012}. 

Figure~\ref{fig:state_evo}b) illustrates the evolution induced by $CU_\textsc{xz}$, when the input states $\ket{H}$ and $\ket{\psi_1}$ are prepared using an entangled resource $\ket\Psi$, rather than directly. The results of the previously discussed distinguishability experiments for this case are shown in Fig.~\ref{fig:results_dist}b). In Fig.~\ref{fig:results_dist}c) they are compared to the case of local preparation and to standard quantum mechanics for a fixed input state and a fixed gate, respectively. Again, consistency of our simulation is ensured by a quantum state fidelity of $\mathcal{F}=0.9996(3)$ between the input and output states of $\rho_\textsc{ctc}$

In our simulation we find that the CTC-system can indeed achieve perfect distinguishability of the (directly prepared) states $\ket{\psi_0}$ and $\ket{\psi_1}$ even for arbitrarily close states if the appropriate gate is implemented, see Fig.~\ref{fig:results_dist}a). Furthermore we show that the advantage over standard quantum mechanics persists for a wide range of non-optimal gate-state combinations, outside of which, however, the CTC-system performs worse (blue points).
Notably, we find that for non-locally prepared input states CTC-assisted state discrimination never performs better than random guessing---a probability of $0.5$---as shown in Fig.~\ref{fig:results_dist}b). The predictions for standard quantum mechanics, in contrast are independent of the way the states $\ket{\psi_0}$ and $\ket{\psi_1}$ are prepared.

$ $\\
\noindent\textbf{Decoherence.}
We further investigated the effect of two important decoherence mechanisms on the simulated CTC-system, shown in Fig.~\ref{fig:setup}a). The first is a single qubit depolarising channel acting on the input state $\ket\psi$, which can be modelled as
\begin{equation}
\rho \mapsto (1-\frac{3p}{4})\rho+\frac{p}{4}\left( \sigma_x \rho \sigma_x+\sigma_y \rho \sigma_y+\sigma_z \rho \sigma_z \right) , 
\label{eq:SingleQubitDecoherence}
\end{equation}
where $(\sigma_x,\sigma_y,\sigma_z)$ are the 3 Pauli matrices and $p\in [0,1]$ quantifies the amount of decoherence. 

The second form of decoherence concerns the controlled unitary $CU_\textsc{xz}$ and is described as
\begin{equation}
\rho \mapsto (1-\ve) CU_\textsc{xz} \ \rho \ CU_\textsc{xz}^\dagger + \ve \rho ,
\label{eq:GateDecoherence}
\end{equation}
where $\ve\in[0,1]$ is the probability of the gate to fail, describing the amount of decoherence that is present. For $\ve=0$ the gate acts as an ideal controlled rotation $CU_\textsc{xz}$, while it performs the identity operation for $\ve=1$. 

\begin{figure}[h!]
\begin{center}
\includegraphics[width=1\columnwidth]{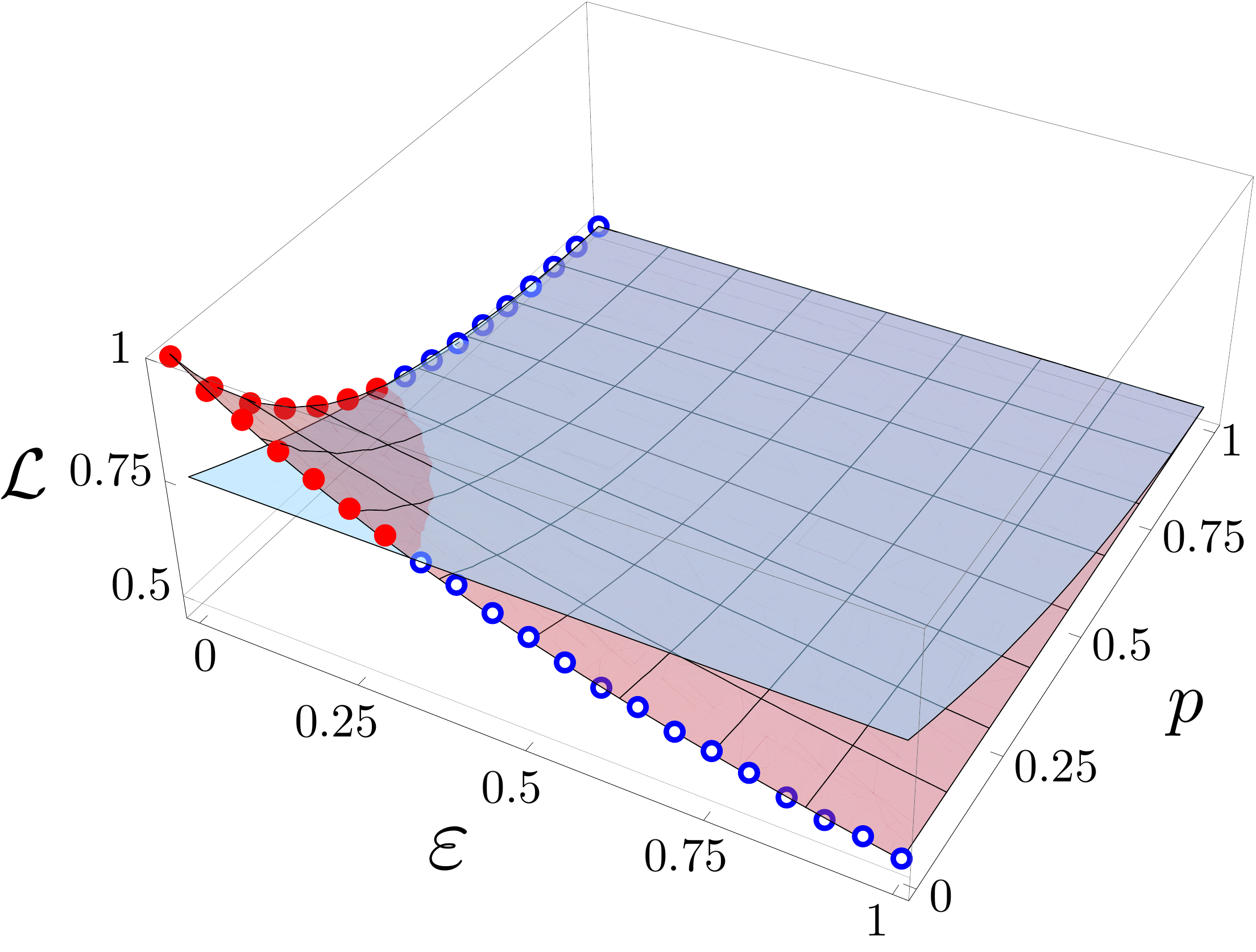}
\end{center}
\vspace{-7mm}
\caption{\textbf{State discrimination as a function of gate and qubit decoherence for locally prepared states.} Here $\ve$ quantifies the decoherence of the unitary interaction $CU_\textsc{xz}$ (with $\theta_\textsc{xz}{=}\pi/4$), which has no analogue in the standard quantum mechanics case and $p$ the single qubit depolarisation of the input qubits $\ket{H}$ and $\ket{\psi_1}$ (with $\phi{=}3\pi/2$). The system demonstrates robustness against both forms of decoherence and the CTC-advantage persists up to $p{=}\sqrt{2}{-}1$ and $\ve{=}\frac 1 3$, respectively. The semi-transparent blue surface represents the optimum in standard quantum mechanics. Error bars obtained from a Monte Carlo routine simulating the Poissonian counting statistics are too small to be visible on the scale of this plot.}
\label{fig:results_dist_decoherence}
\end{figure}

We tested the robustness of the state-discrimination circuit in Fig.~\ref{fig:setup}b)(ii) against both forms of decoherence. For this test we chose $CU_\textsc{xz}$ as a controlled Hadamard (i.e.\ $\theta_\textsc{xz}{=}\pi/4$) and the initial states $\ket{\psi_0}{=}\ket{H}$ and $\ket{\psi_1}{=}\frac{1}{\sqrt{2}}\left(\ket{H}{-}\ket{V}\right)$ (i.e.\ $\phi{=}3\pi/2$). Figure~\ref{fig:results_dist_decoherence} shows the distinguishability $\mathcal{L}$ of the evolved states as a function of both decoherence mechanisms over the whole range of parameters $p\in[0,1]$ and $\ve\in [0,1]$. Note, that the decoherence channel in Eq.~\eqref{eq:GateDecoherence} does not have an analogue in the standard quantum mechanics case (i.e.\ without a CTC), hence only the channel in Eq.~\eqref{eq:SingleQubitDecoherence} is considered for comparison. It is further naturally assumed, that the experimenter has no knowledge of the specific details of the decoherence and therefore implements the optimal measurements for the decoherence-free case. The physical validity of the simulation is ensured by consistency of $\rho_{\textsc{ctc}}$ across the boundary of the wormhole with an average fidelity of $\mathcal{F}=0.997(4)$.

It is worth noting, that the interpretation of decoherence effects in the circuit in Fig.~\ref{fig:setup}a) is very different from the linear scenario without a CTC. In the case of single-qubit depolarisation the initially pure input state becomes mixed. In contrast to the linear case now an important distinction has to be made with respect to the origin of the decoherence. If it results from an interaction with the environment, which is the case considered here, then $\rho_\textsc{ctc}$ ``sees'' an improper mixture and adjusts to the mixed density matrix of the input state. If, however, the origin of the mixture is classical fluctuations in the preparation apparatus, then shot-by-shot pure states enter the circuit and the consistency relation holds accordingly shot-by-shot, resulting in a proper mixture at the output. This shows that in the presence of a CTC it would be possible to identify the origin of the decoherence in an experimental setup.

Furthermore, careful analysis of the decoherence of the unitary gate $U$ reveals parallels to effects seen in non-local state preparation. The decoherence is assumed to arise from non-local coupling to the environment. Again, due to a lack of classical knowledge of the outcome of an eventual measurement of the environment, $\rho_{\textsc{ctc}}$ ``sees'' the mixed process in Eq.~\eqref{eq:GateDecoherence} in every run of the experiment.
In the case of full decoherence the distinguishability is reduced to $0.5$ as in standard quantum mechanics. The differences between local and non-local decoherence in their interpretation and effect is one of the key insights from our simulation.

\section{Discussion}
Quantum simulation is a versatile and powerful tool for investigating quantum systems that are hard or even impossible to access in practice~\cite{Casanova2011}. Although no CTCs have been discovered to date, quantum simulation nonetheless enables us to study their unique properties and behaviour. Here we simulated the immediate adaption of $\rho_\textsc{ctc}$ to changes in the CTC's environment and in particular the effect of different forms of decoherence. We also show that the non-linearity inherent in the system is in fact not uniform as shown in Fig.~\ref{fig:results_nonlinear_bacon}, suggesting that non-linear effects only become apparent in certain scenarios and for a specific set of measurements.  

Moreover, we find intriguing differences with respect to nominally equivalent ways of pure state preparation. Although acknowledged in Ref.~\cite{Cavalcanti2012} this feature has not been further investigated in the present literature. Importantly this effect arises due to consistency with relativity, in contrast to the similar effect for mixed quantum states discussed earlier, which is a direct result of the non-linearity of the system~\cite{ralph2010information}. 

Our study of the Deutsch model provides insights into the role of causal structures and non-linearities in quantum mechanics, which is essential for an eventual reconciliation with general relativity.
\subsection*{Acknowledgments}
We thank Nathan Walk and Nicolas Menicucci for insightful discussions. We acknowledge financial support from the ARC Centres of Excellence for Engineered Quantum Systems (CE110001013) and Quantum Computation and Communication Technology (CE110001027). A.G.W.\ and T.C.R.\ acknowledge support from a UQ Vice-Chancellor's Senior Research Fellowship.
\subsection*{Author contributions}
M.R., M.A.B., C.R.M.\ and T.C.R.\ developed the concepts, designed the experiment, analysed the results and wrote the paper. M.R.\ performed the experiments and analysed data. T.C.R.\ and A.G.W.\ supervised the project and edited the manuscript.

\onecolumngrid
\renewcommand{\theequation}{S\arabic{equation}}
\renewcommand{\thefigure}{S\arabic{figure}}
\renewcommand{\thetable}{\Roman{table}}
\renewcommand{\thesection}{S\Roman{section}}
\setcounter{equation}{0}
\setcounter{figure}{0}
\begin{center}
{\bf \large Supplemental Material}
\end{center}
\section{Distinguishing (mixed) quantum states}
\label{Sec:Supp1}

\noindent The measure $\mathcal L$ introduced in Eq.~\eqref{eq:sigmaZ} has an operational interpretation as the probability of obtaining the outcome ``different'' when comparing two quantum states by means of a single projective measurement on each system. Notably, the minimum-error measurement for discriminating two quantum states is indeed a projective measurement in a direction that depends on the two states~\cite{Helstrom1969,Jaeger1995,Herzog2004}. Hence, considering only projective measurements is not a restriction and the measure is optimal with the right choice of measurement direction. This result in particular also holds for mixed quantum states, which will become very relevant in the next section.\\
\\
The situation in the main text can be recast as a game where Alice prepares two quantum systems, one in state $\ket{\psi_0}$ and one in state $\ket{\psi_1}$ and sends them to Bob, whose task is to determine whether they are different or not. If Alice indeed sends two different states, then the measure $\mathcal L$ is understood as Bob's probability of either guessing both states correctly or both incorrectly, which are the two cases where he successfully distinguishes the states. Hence, $\mathcal L$ is a natural measure for this task and---given that Bob uses the knowledge about the states to be distinguished---also optimal. In fact, given an optimal choice of measurement direction, $\mathcal L$ is directly related to the trace-distance metric $\mathcal D$ and therefore a similarly suitable measure of distinguishability:
\begin{equation*}
\mathcal L =  p_{\text{correct}}^2 + p_{\text{error}}^2 = 1 - \frac 1 2 |\braket{\psi_0}{\psi_1}|^2 = \frac 1 2 \bigl(1 + \mathcal{D}^2 \bigr)
\label{eq:SuppOptL}
\end{equation*}

\subsection{Optimal CTC implementation}
\label{Sec:Supp2}
\begin{figure}[h!]
\begin{center}
\includegraphics[width=1\textwidth]{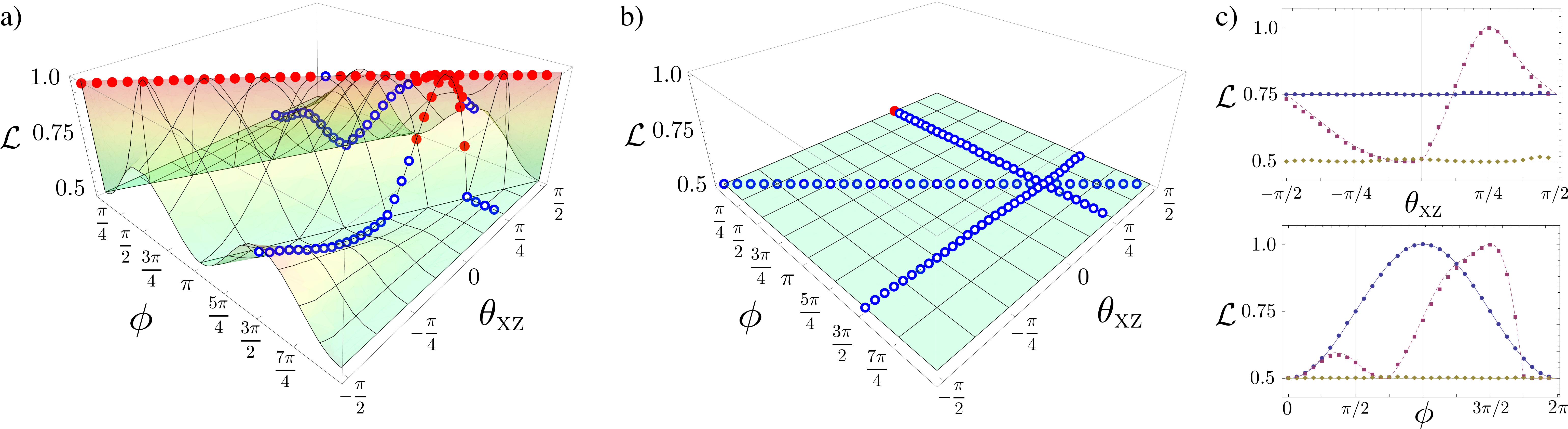}
\end{center}
\vspace{-5mm}
\caption{\textbf{Experimental results for the state discrimination scenario with an optimal CTC implementation.} Probability of state discrimination for \textbf{a)} locally prepared and \textbf{b)} non-locally prepared states $\ket{\psi_0}{=}\ket{H}$ and $\ket{\psi_1}=\cos (\frac\phi 2)\ket{H}+\sin(\frac\phi 2)\ket{V}$ as measured by $\mathcal{L}$. The surface represents the theoretically predicted probability for the optimal CTC implementation, depending on the state and gate parameters $\phi$ and $\theta_\textsc{xz}$, respectively. Solid, red (open, blue) data-points indicate better (worse) performance than standard quantum mechanics, which also implements the optimal measurement, making use of all the available information. \textbf{c)} Cross-sectional views of the combined plots a) and b) as in Fig.~\ref{fig:results_dist}. Error bars obtained from a Monte Carlo routine simulating the Poissonian counting statistics are too small to be visible on the scale of this plot.}
\label{fig:SuppResultsFullOpt}
\end{figure}

\noindent In the main text we investigated the case where the controlled unitary $CU_{\textsc{xz}}$ is chosen non-optimally for the state $\ket{\psi_1}$. Notably, this is considered a conscious choice of the experimenter, in contrast to decoherence of the gate, which is beyond their control. Hence, the knowledge about $\theta_{\textsc{xz}}$ is available and can be used to optimize the measurement direction of the final projective measurement as is done in the case of standard quantum mechanics. Although in the non-optimal CTC-case the output states are mixed, this does not change the fact that the optimal measurement is projective. Hence, the quantity $\mathcal L$ can be optimized depending on the states $\ket{\psi_0}, \ket{\psi_1}$ and the gate $CU_{\textsc{xz}}$. The corresponding results are shown in Fig.~\ref{fig:SuppResultsFullOpt}, which differs from Fig.~\ref{fig:results_dist} in that the CTC case is now optimal for the chosen gate.

Note that now the CTC-circuit always achieves the best distinguishability of $1/2$ for non-local state preparation. In the local case the advantage over standard quantum mechanics is extended to a wider range of non-optimal combinations. Furthermore, we observe recovery of distinguishability for combinations far from optimal, see Fig.~\ref{fig:SuppResultsFullOpt}.

\subsection{State identification}
\label{Sec:Supp3}
\begin{figure}[h!]
\begin{center}
\includegraphics[width=1\textwidth]{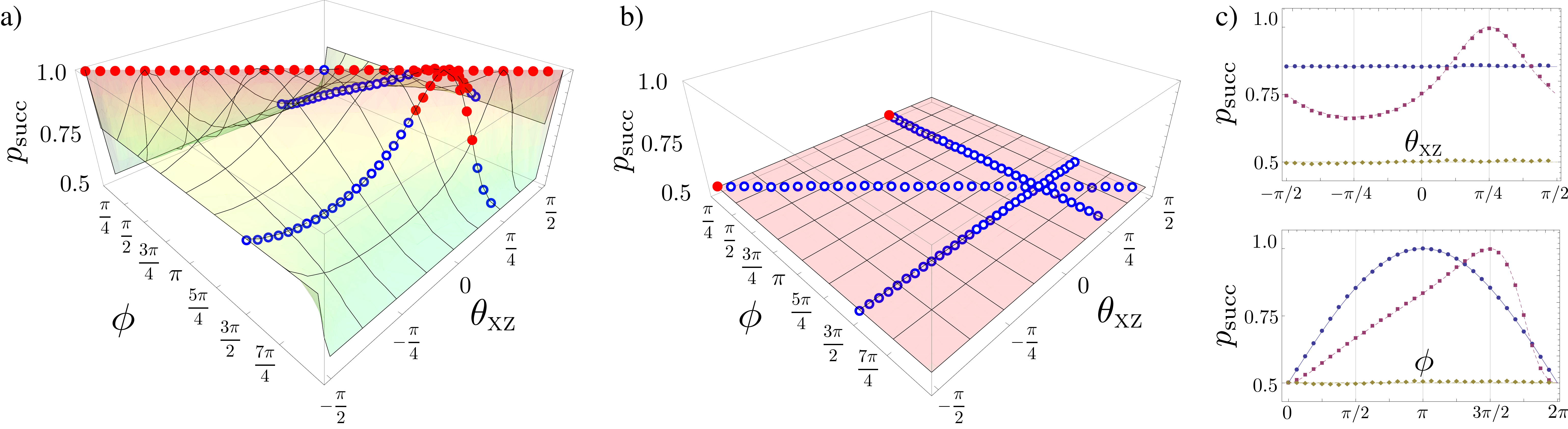}
\end{center}
\vspace{-5mm}
\caption{\textbf{Experimental results for the state identification scenario.} Probability of state identification $p_{\text{succ}}$ for \textbf{a)} locally prepared and \textbf{b)} non-locally prepared states $\ket{\psi_0}{=}\ket{H}$ and $\ket{\psi_1}=\cos (\frac\phi 2)\ket{H}+\sin(\frac\phi 2)\ket{V}$. The surface represents the theoretically predicted probability depending on the state and gate parameters $\phi$ and $\theta_\textsc{xz}$, respectively. Solid, red (open, blue) data-points indicate better (worse) performance than standard quantum mechanics, which implements the optimal measurement, making use of all the available information. \textbf{c)} Cross-sectional views of the combined plots a) and b) as in Fig.~\ref{fig:results_dist}. Error bars obtained from a Monte Carlo routine simulating the Poissonian counting statistics are too small to be visible on the scale of this plot.}
\label{fig:SuppResultsIdentification}
\end{figure}

\noindent As an alternative approach we consider a scenario where Alice prepares two known quantum states $\ket{\psi_0}$ and $\ket{\psi_1}$ at random and sends them---one at a time---to Bob, who is given the task of identifying each of the states. Similarly to the state-discrimination case, the optimal measurement is a projective measurement in a direction that depends on the two states. The figure of merit that is intrinsically related to this task is the probability of success,
\begin{equation*}
p_{\text{succ}}= p_{\ket{\psi_0}} p(\psi_0\mid\ket{\psi_0}) + p_{\ket{\psi_1}} p(\psi_1\mid\ket{\psi_1}) ,
\label{eq:SuppProbSuccess}
\end{equation*}
where $p_{\ket{\psi}}$ is the probability for the state $\ket{\psi}$ to be sent and $p(\phi\mid\ket{\psi})$ is Bob's probability for guessing $\ket{\phi}$ in the case where he received the state $\ket{\psi}$. The optimal measurement direction can again be chosen based on knowledge of the two states to be identified. In the scenario considered here, this information is available to Bob and the states are prepared with equal probability. The optimal probability of success is then given by
\begin{equation*}
p_{\text{succ}}=\frac 1 2 \bigl(1+\sqrt{1-|\braket{\psi_0}{\psi_1}|^2} \bigr) = \frac 1 2 \bigl(1 + \mathcal D) .
\label{eq:SuppOptProbSuccess}
\end{equation*}
Hence, in this case $p_{\text{succ}}$ is also directly related to $\mathcal D$, making it an equivalent measure of distinguishability. We have analyzed our experiment from this point of view and find the same qualitative behavior: the CTC circuit outperforms standard quantum mechanics for the optimally chosen unitary interaction, as well as for a range of non-optimal choices. The results are shown in Fig.~\ref{fig:SuppResultsIdentification}, which parallels Fig.~\ref{fig:results_dist}.

Note that in contrast to Fig.~\ref{fig:results_dist}, the CTC-circuit always achieves, but never surpasses $1/2$ in the case of non-locally prepared states.

\renewcommand\bibsection{\subsection{\refname}}

\end{document}